\documentclass[12pt,pra,aps,amssymb,amsfonts,amsmath,tightenlines]{revtex4}
\usepackage{dsfont}
\usepackage[dvips]{graphicx}
\usepackage{amssymb,amsfonts}

\newcommand{\half}{\mbox{$\textstyle \frac{1}{2}$}}

\newcommand{\re}{\mbox{$\rm e$}}

\newcommand{\rd}{\mbox{$\rm d$}}

\begin{document}

\title{Information of Interest}

\author{Dorje~C.~Brody${}^1$
and Robyn~L.~Friedman${}^{1,2}$ }

\affiliation{${}^1$ Department of Mathematics, Imperial College
London, London SW7 2BZ, UK \\ ${}^2$Royal Bank of Scotland, 135
Bishopsgate, London EC2M 3UR, UK }


\begin{abstract}
A pricing formula for discount bonds, based on the consideration of
the market perception of future liquidity risk, is established. An
information-based model for liquidity is then introduced, which is
used to obtain an expression for the bond price. Analysis of the
bond price dynamics shows that the bond volatility is determined by
prices of certain weighted perpetual annuities. Pricing formulae for
interest rate derivatives are derived.
\\ \vspace{0.1cm}

\noindent Working paper. This version: {\today}.

\noindent Email: dorje@imperial.ac.uk, robyn.freidman@rbs.com
\end{abstract}


\maketitle



\textbf{1. Introduction}. We would like to present here an idea
concerning how to generate interest rate dynamics from elementary
economic considerations. There are of course numerous economic
factors that affect the movement of interest rates, and causal
relations that hold between these factors are often difficult to
disentangle. Hence, rather than attempting to address a range of
factors simultaneously, we will focus on one key factor that appears
important in determining the interest rate term structure; namely,
the liquidity risk, in the narrow sense of cash demand. Our
objective is to build an information-based model that reflects the
market perception of future liquidity risk, and use it for the
pricing and general risk management of interest rate derivatives.

In the framework of arbitrage-free pricing theory the value of an
asset is determined by the risk-adjusted expectation of the suitably
discounted cash flow. Thus in principle one can \textit{derive} the
price process of an asset by the specification of the random cash
flow, along with the market filtration and the discounting factor.
Such an idea has been applied successfully to obtain the price
process of, for example, credit-risky bonds (Brody \textit{et al}.
2007) or reinsurance-related products (Brody \textit{et al}. 2008).
When it comes to the modelling of interest rate term structure,
however, the matter is made somewhat more complicated, because the
cash flow of a discount bond is not random, and thus one has to
specify the discount factor to deduce the bond price. One is then
led to the specification of the short rate or forward rate
processes, but such an approach is undesirable if the objective is
to generate from the outset the dynamics of the term structure. This
forces us to take an alternative route for deriving bond prices and
the associated rates.

The new pricing framework outlined here, based on the consideration
of liquidity risk, has several advantages worth noting: (i) economic
interpretations of the discount function, the associated rates, and
their volatilities become intuitive; (ii) initial term structure can
be specified exogenously in a straightforward manner; (iii)
arbitrage-free dynamics of interest rates emerge endogenously in
such a manner that they are consistent with the market perception of
future cash demand; (iv) semi-analytic formulae for caplet and
swaption, expressed in terms of elementary Gaussian integrations,
can be obtained; and (v) risk premium can in principle be estimated
from prices of interest rate derivatives. To illustrate the role
played by liquidity risk in determining interest rate systems, let
us begin by examining deterministic term structures.

\vspace{0.15cm}

\textbf{2. Deterministic term structures}. Consider first the
initial discount function $P_{0T}$. It should be evident that
positivity of nominal rates implies that the discount function
$P_{0T}$ is decreasing in the maturity variable $T$. Furthermore, a
common sense argument shows that a bond with infinite maturity has
no value. Thus $P_{0T}$ can be thought of as defining a right-side
cumulative distribution function on the positive real line ${\mathds
R}_+$. In particular, $\rho_0(T) = -\partial_T P_{0T}$ defines a
density function over ${\mathds R}_+$. Put the matter differently,
the positive interest term structure implies the existence of a
random variable $X$ on a probability space with measure ${\mathbb
Q}$ such that we have $P_{0T}={\mathbb Q}(X\geq T)$. Based on this
observation a general arbitrage-free dynamical equation satisfied by
the term structure density process $\rho_t(T) = -\partial_T P_{tT}$
was obtained by Brody \& Hughston (2001). In the present
investigation, however, we would like to pay more attention to the
interpretation of the cumulative distribution $P_{0T}$, the meaning
of the random variable $X$, and the role of the probability measure
${\mathbb Q}$, in such a way that new interest rate models can be
created.

We remark that it is reasonable to regard the random variable $X$ as
representing the occurrence time of future liquidity issues, at
least to first approximation. From the viewpoint of the buyer of a
bond, if with high probability cash is needed before the maturity
$T$, then purchase will be made only if the bond price is
sufficiently low. Likewise, the seller of a bond would be willing to
pay a high premium if there is a likely need for cash before time
$T$. Thus $P_{0T}$ represents a survival function, where `survival'
means lack of liquidity crisis. It is worth noting that the
interplay between liquidity and interest rate has long been
discussed in the economics literature. To this end we refer to the
presidential address delivered at the eightieth annual meeting of
the American Economic Association (Friedman 1968) for further
insights.

What we would like to establish here is the fact that the price of a
discount bond with maturity $T$ is determined by the risk-adjusted
probability that the liquidity crisis arises beyond time $T$:
$P_{0T}={\mathbb Q}(X\geq T)$. That ${\mathbb Q}$ represents the
risk-neutral measure will be shown later, but let us for the moment
assume that this is the case. Then the risk-neutral hazard rate
associated with liquidity crisis is just the initial forward rate
$f_{0T}$. Therefore, for a small $\rd T$ we have
\begin{eqnarray}
f_{0T}\,\rd T = {\mathbb Q}\left( X\in[T,T+\rd T]\, |\, X\geq T
\right).
\end{eqnarray}
In other words, $f_{0T}\,\rd T$ is the \textit{a priori}
risk-neutral probability of a liquidity crisis occurring in an
infinitesimal interval $[T,T+\rd T]$, conditional upon survival
until time $T$.

More generally, in the case of a deterministic interest rate term
structure, the price $P_{tT}$ at time $t$ of a bond that matures at
$T$ is given by the risk-neutral probability of survival until $T$
conditional on survival until $t$:
\begin{eqnarray}
P_{tT} = {\mathbb Q}\left( X\geq T\, |\, X\geq t \right).
\label{eq:2}
\end{eqnarray}
This can be verified by use of the Bayes formula, which shows that
the right side of (\ref{eq:2}) is given by $P_{0T}/P_{0t}$. But this
is just the bond price $P_{tT}$ in the case of a deterministic term
structure. Thus in a deterministic interest rate system we can
calibrate the initial term structure density $\rho_0(T)$ using the
initial yield curve, from which the subsequent evolution is
determined in accordance with (\ref{eq:2}).

\vspace{0.15cm}

\textbf{3. Market information about future liquidity}. Our aim now
is to extend the deterministic model (\ref{eq:2}) into a dynamical
one without losing the key economic interpretation. That is to say,
we would like to retain the fact that the bond price represents the
conditional risk-neutral probability that the liquidity issue arises
beyond time $T$. The problem therefore is to identify the relevant
conditioning. In the case of a deterministic term structure
(\ref{eq:2}) the conditioning is given merely by the event $X\geq
t$. In a dynamical setup, however, market participants accumulate
noisy information concerning future liquidity risk. It is this noisy
observation of the timing $X$ of the future cash demand that
generates random movements in the bond price. Thus if we let $\{
{\mathcal F}_t\}$ denote the information generated by this
observation, then the price of a discount bond is given by the
conditional probability
\begin{eqnarray}
P_{tT} = {\mathbb Q}\left( X\geq T\, |\, (X\geq t) \cap {\mathcal
F}_t \right). \label{eq:3}
\end{eqnarray}
Evidently, the random variable representing the timing of cash
demand itself may change in time. In the present investigation,
however, we shall confine our analysis to models based on fixed
$X$.

If we apply the Bayes formula, then (\ref{eq:3}) can be expressed
in a more intuitive form
\begin{eqnarray}
P_{tT} = \frac{{\mathbb Q}\left( X\geq T | {\mathcal F}_t \right)}
{{\mathbb Q}\left( X\geq t | {\mathcal F}_t \right)}. \label{eq:4}
\end{eqnarray}
This is the pricing formula for a discount bond that we propose
here. In obtaining (\ref{eq:4}) we have made use of the fact that
$(X\geq T) \cap (X\geq t) = (X\geq T)$.

\vspace{0.15cm}

\textbf{4. An elementary model for bond price}. To proceed we
introduce a specific model for $\{ {\mathcal F}_t\}$. Since in the
present formulation what concerns market participants is the value
of $X$, the `signal' component of the observation must be generated
in some form by $X$ itself. In addition, there is an independent
noise that obscures the value of $X$. Motivated by the approach
introduced in Macrina (2006) and in Brody \textit{et al}. (2007) for
the information-based asset pricing framework, let us consider a
simple model whereby the information concerning the value of $X$ is
revealed to the market linearly in time at a constant rate $\sigma$,
and the noise is generated by an independent Brownian motion
$\{B_t\}$, defined on a probability space with measure ${\mathbb
Q}$. Thus the information generating process is given by
\begin{eqnarray}
\xi_t = \sigma t \phi(X) + B_t , \label{eq:5}
\end{eqnarray}
where $\phi(x)$ is a smooth invertible function. In other words, we
assume that the filtration ${\mathcal F}_t$ is given by the sigma
algebra generated by $\{\xi_s\}_{0\leq s\leq t}$. Note that a more
coherent formulation is obtained if we replace the Brownian noise
$B_t$ by a `killed' Brownian noise ${\mathds 1}\{X\geq t\}B_t$.
However, since we are interested in events on $\{X\geq t\}$, and
since this alternation does not affect the bond pricing formula, we
shall be using (\ref{eq:5}) for simplicity of exposition. As regards
the choice of the function $\phi(x)$ we shall have more to say
shortly, but let us for the moment proceed with generality.

We note that since the magnitude of the signal-to-noise ratio is
given by $\sigma\sqrt{t}$, the value of $X$ will be revealed
asymptotically, that is, $X$ is ${\mathcal F}_\infty$-measurable.
Along with the fact that $\{\xi_t\}$ of (\ref{eq:5}) is Markovian,
we find that the bond pricing formula simplifies in this model to
\begin{eqnarray}
P_{tT} = \frac{{\mathbb Q}\left( X\geq T | \xi_t \right)} {{\mathbb
Q}\left( X\geq t | \xi_t \right)}. \label{eq:6}
\end{eqnarray}
For the calculation of the bond price (\ref{eq:6}) we consider the
following joint probability ${\mathbb Q}\left( (X\geq T) \cap (\xi_t
\in \rd\xi) \right)$. Then, on account of the definition
(\ref{eq:5}), we have
\begin{eqnarray}
{\mathbb Q}\left( (X\geq T) \cap (\xi_t \in \rd\xi) \right) =
\int_T^\infty {\mathbb Q}\left(B_t\in[\rd\xi-\sigma t\phi(x)]
\right) \rho_0(x) \rd x, \label{eq:7}
\end{eqnarray}
where $\rho_0(x)=-\partial_x P_{0x}$ is the initial term structure
density. By substituting the density function for the Brownian
motion we obtain the following expression:
\begin{eqnarray}
P_{tT} = \frac{ \int_T^\infty \rho_0(x)\, \re^{\sigma \phi(x) \xi_t
- \frac{1}{2}\sigma^2 \phi^2(x) t}\rd x} { \int_t^\infty \rho_0(x)\,
\re^{\sigma \phi(x) \xi_t - \frac{1}{2}\sigma^2 \phi^2(x) t}\rd x}.
\label{eq:8}
\end{eqnarray}
It is interesting to observe that the form of the bond price thus
obtained is closely related to the general positive interest
representation obtained by Flesaker \& Hughston (1996).

\begin{figure}
\begin{center}\vspace{-0.0cm}
  \includegraphics[scale=0.7]{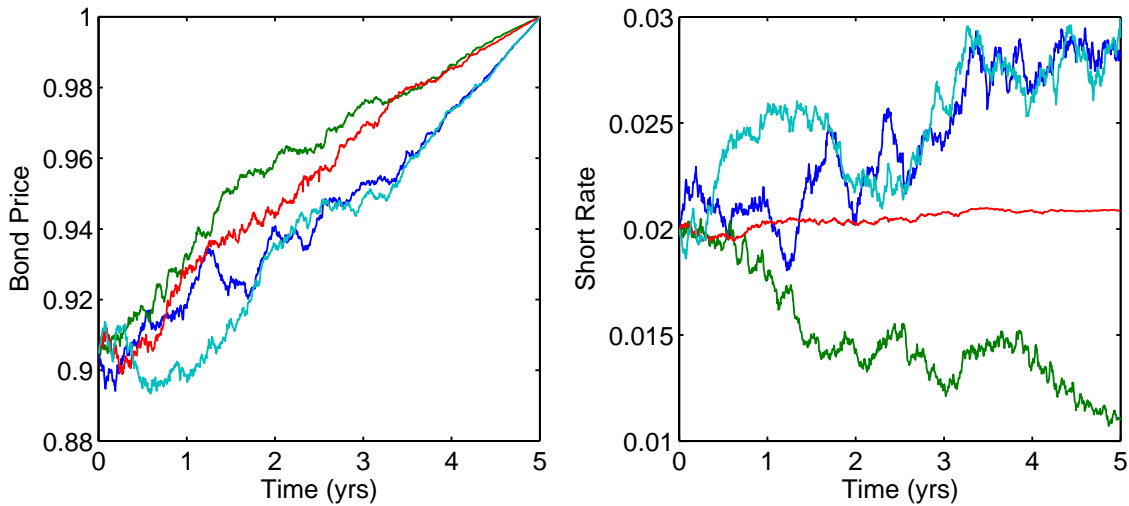}
  \vspace{-0.0cm}
  \caption{Sample paths of the discount function (\ref{eq:8}) and
  the associated short rate (\ref{eq:9}). The information-adjusting
  function is set as $\phi(x)=\re^{-0.025x}$, and the initial term
  structure is assumed flat so that $P_{0T}=\re^{-0.02T}$. The
  information flow rate is set as $\sigma=0.3$, and the bond maturity
  is 5 years.
  \label{fig:1}
  }
\end{center}
\end{figure}

As in the deterministic case, the model can be calibrated exactly
against the initial yield curve according to the prescription
$\rho_0(x)=-\partial_x P_{0x}$. The subsequent evolution is then
determined by the Markovian market information process. In this
respect the model has a feature resembling the Markov-functional
models (Hunt \& Kennedy 2000). The remaining degree of freedom,
namely, the parameter $\sigma$, can be calibrated by use of
derivative prices. This will be discussed later.

From the bond price (\ref{eq:8}) we can infer the implied short rate
$r_t=-\partial_T P_{tT}|_{T=t}$. This is given by
\begin{eqnarray}
r_t = \frac{ \rho_0(t)\, \re^{\sigma \phi(t) \xi_t-\frac{1}{2}
\sigma^2 \phi^2(t)t} }{\int_t^\infty \rho_0(x)\, \re^{\sigma \phi(x)
\xi_t - \frac{1}{2}\sigma^2 \phi^2(x) t}\rd x}. \label{eq:9}
\end{eqnarray}
We draw attention to the observation made in Brody \& Hughston
(2001) that the short rate is the negative expectation of the
differential operator $\partial_X=\partial/\partial X$ defined by
the action $\partial_X\psi(x)=\partial_x\psi(x)$ on any test
function $\psi(x)$. In the present context, this means that
formally we can write $r_t=-{\mathbb E}[\partial_X|(X\geq t)\cap
\xi_t]$. Here and in what follows expectations are taken with
respect to the ${\mathbb Q}$ measure unless otherwise specified.
The instantaneous forward rate $f_{tT}=-\partial_T \ln P_{tT}$ is
expressed analogously as
\begin{eqnarray}
f_{tT} = \frac{ \rho_0(T)\, \re^{\sigma \phi(T) \xi_t-\frac{1}{2}
\sigma^2 \phi^2(T) t}}{\int_T^\infty \rho_0(x)\, \re^{\sigma \phi(x)
\xi_t - \frac{1}{2}\sigma^2 \phi^2(x) t}\rd x}. \label{eq:10}
\end{eqnarray}

\vspace{0.1cm}

\textit{Example}. Consider a flat initial term structure given by
$P_{0T}=\re^{-rT}$. The associated \textit{a priori} density
function is then exponential: $\rho_0(T)=r\re^{-rT}$. In a
\textit{linear information model}, we have $\phi(x)=x$. Substitution
of these in (\ref{eq:8}) yields the following bond price process
\begin{eqnarray}
P_{tT} = \frac{N\left(\frac{\xi_t-r/\sigma}{\sqrt{t}}-\sigma T
\sqrt{t}\right)}{N\left(\frac{\xi_t-r/\sigma}{\sqrt{t}}-\sigma
t\sqrt{t} \right)}, \label{eq:11}
\end{eqnarray}
where $N(x)$ is the normal distribution function.

\vspace{0.15cm}

\textbf{5. Dynamics of the bond price}. We now turn to the analysis
of the discount bond dynamics. We find it convenient to introduce
the following one-parameter family of processes
\begin{eqnarray}
{\hat\Phi}_{tu} = \frac{ \int_u^\infty \phi(x) \rho_0(x)\,
\re^{\sigma \phi(x) \xi_t - \frac{1}{2}\sigma^2 \phi^2(x) t}\rd x}
{\int_u^\infty \rho_0(x)\, \re^{\sigma \phi(x) \xi_t - \frac{1}{2}
\sigma^2 \phi^2(x) t}\rd x}. \label{eq:12}
\end{eqnarray}
For $u=t$ this corresponds to the conditional expectation ${\hat
\Phi}_{tt}= {\mathbb E}[\phi(X){\mathds 1}\{X\geq t\}| (X\geq t)
\cap \xi_t]$. That ${\hat\Phi}_{tt}$ represents the said conditional
expectation can be seen from (\ref{eq:3}) and (\ref{eq:8}) which
allows us to read off the conditional probability law for $X$.

The bond price dynamics can be deduced by taking the stochastic
differential of (\ref{eq:8}). A calculation shows that
\begin{eqnarray}
\frac{\rd P_{tT}}{P_{tT}} = r_t \rd t + \sigma \Sigma_{tT} \rd W_t.
\label{eq:13}
\end{eqnarray}
Here we have defined
\begin{eqnarray}
\Sigma_{tT} = {\hat\Phi}_{tT} - {\hat\Phi}_{tt} \label{eq:14}
\end{eqnarray}
and
\begin{eqnarray}
W_t = \xi_t - \sigma \int_0^t {\hat\Phi}_{ss} \rd s. \label{eq:15}
\end{eqnarray}
The key result that we shall establish below is the fact that
$\{W_t\}$ thus defined is a ${\mathbb Q}$-Brownian motion on
$\{X\geq t\}$ with respect to the market filtration $\{{\mathcal
G}_t\}$ determined jointly by $\{{\mathcal F}_t\}$ and the sigma
algebra generated by $\{X\geq s\}_{0\leq s\leq t}$. (More precisely,
the process defined by $\{{\mathds 1}\{X\geq t\}W_t\}$ is the killed
Brownian motion.) It follows from (\ref{eq:13}) that the probability
measure ${\mathbb Q}$ can be identified with the risk-neutral
measure, since the drift of the bond in this measure is given by the
short rate. Following the terminology of Wiener we shall refer to
$\{W_t\}$ as the innovations process, because $\{W_t\}$ measures the
arrival of new information to the market concerning future liquidity
risk. The dynamical equation (\ref{eq:13}) also shows that the model
under consideration is in fact of a single-factor diffusion type,
with a (hedgeable) stochastic volatility and stochastic rates.

We observe from (\ref{eq:14}) that the bond volatility is determined
by the difference between forward conditional expectation ${\hat
\Phi}_{tT}= P_{tT}^{-1}{\mathbb E}[\phi(X){\mathds 1}\{X\geq T\} |
(X\geq t) \cap \xi_t]$ of $\phi(X)$ to time $T$ and the conditional
expectation ${\hat\Phi}_{tt}= {\mathbb E}[\phi(X){\mathds 1}\{X\geq
t\}| (X\geq t) \cap \xi_t]$ of $\phi(X)$ at time $t$. These
expectations are related to the concept of advanced and backward
transforms considered in survival analysis (Efron \& Johnstone
1990). Therefore, if the forward expectation of the function
$\phi(X)$ of the timing for future cash demand is close to the
current (time $t$) expectation, then the bond volatility is low.
Conversely, if there is a large discrepancy between the forward
expectation and the current expectation of $\phi(X)$, then the bond
price process becomes volatile.

\vspace{0.1cm}

\textit{Example}. In the case of a linear information model
$\phi(x)=x$, we can, in fact, assign a more direct financial
interpretation to the meaning of the bond volatility, by virtue of
an observation made in Brody \& Hughston (2001) that the expectation
${\mathbb E}[X]$ of the random variable $X$ is the initial price of
the perpetual annuity. In the present framework, $\{{\hat
\Phi}_{tt}\}$ for $\phi(x)=x$ represents the shifted price process
of the annuity. Specifically, we have, on account of integration by
parts using the relation $\rho_t(x)= -\partial_x P_{tx}$, the
following representation
\begin{eqnarray}
{\hat\Phi}_{tt} = t + \int_t^\infty P_{tx}\rd x.  \label{eq:16}
\end{eqnarray}
On the other hand, $\{{\hat\Phi}_{tT}\}$ can be thought of as its
forward price in the sense that
\begin{eqnarray}
{\hat\Phi}_{tT} = T + P_{tT}^{-1}\int_T^\infty P_{tx}\rd x.
\label{eq:17}
\end{eqnarray}
Hence the bond volatility, when $\phi(x)=x$, is given by the
difference between the forward and current prices of the perpetual
annuity plus the time gap $T-t$. In this way we are able to identify
an elementary economic interpretation for the bond price dynamics.
Furthermore, it also implies that volatility-related products for
discount bonds are essentially exotic derivatives on annuities in
this model.

\vspace{0.1cm}

To show that the innovations process $\{W_t\}$ is a Brownian motion
on $\{X\geq t\}$, that is, $\{{\mathds 1}\{X\geq t\}W_t\}$ is the
killed Brownian motion, we note that since $(\rd W_t)^2=\rd t$ it
suffices to verify that $\{W_t\}$ is a martingale. The proof can be
sketched as follows. Writing ${\mathbb E}_t[-]$ for the conditional
expectation with respect to $\{{\mathcal G}_t\}$ and restricting
attention on $\{X\geq t\}$ we obtain
\begin{eqnarray}
{\mathbb E}_t[W_u] &=& {\mathbb E}_t[\xi_u] - \sigma \int_0^u
{\mathbb E}_t[{\hat\Phi}_{ss}]\rd s \nonumber \\ &=& \sigma u
{\mathbb E}_t[\phi(X)] + {\mathbb E}_t[B_u] - \sigma \int_0^t
{\hat\Phi}_{ss}\rd s - \sigma \int_t^u {\mathbb
E}_t[{\hat\Phi}_{ss}]\rd s . \label{eq:18}
\end{eqnarray}
We now observe the fact that ${\mathds 1}\{X \geq t\}{\mathbb E}_t
[\phi(X)] = {\mathds 1}\{X\geq t\} {\hat\Phi}_{tt}$ (cf. Bielecki \&
Rutkowski 2002, chapter 5), which shows that on $\{X\geq t\}$ the
random variable ${\hat\Phi}_{tt}$ is the conditional expectation of
$\phi(X)$ with respect to $\{{\mathcal G}_t\}$. It follows that on
$\{X\geq t\}$ we have
\begin{eqnarray}
{\mathbb E}_t[W_u] = \sigma t {\mathbb E}_t[\phi(X)] + {\mathbb
E}_t[B_u] - \sigma \int_0^t {\hat\Phi}_{ss}\rd s. \label{eq:18.5}
\end{eqnarray}
Now from the tower property of conditional expectation we find
${\mathbb E}_t[B_u]={\mathbb E}_t[{\mathbb E}[B_u|{\mathcal
F}_t^B,X]]={\mathbb E}_t[B_t]$, and since ${\mathbb E}_t[\xi_t]=
\xi_t$ we deduce the martingale condition ${\mathbb E}_t[W_u]= W_t$.
It follows on account of L\'evy's characterisation that $\{W_t\}$ is
a ${\mathbb Q}$-Brownian motion on $\{X\geq t\}$.

In the event $\{X< t\}$ the bond price goes to zero and the dynamics
is terminated, resulting in the killing of the Brownian motion. Such
a hypothetical event corresponds to the `quenching' of the market
where liquidity has completely dried out and there is no
transferrable fund available.

We note that in terms of the risk-neutral Brownian motion
$\{W_t\}$ the forward rate dynamics can be expressed manifestly in
the HJM form:
\begin{eqnarray}
\rd f_{tT} = \sigma^2 \Sigma_{tT}(\partial_T\Sigma_{tT})\rd t -
\sigma(\partial_T\Sigma_{tT})\rd W_t.  \label{eq:19}
\end{eqnarray}
This follows from taking the stochastic differential of
(\ref{eq:10}), and making use of expressions (\ref{eq:12}),
(\ref{eq:14}), and (\ref{eq:15}). A calculation shows that the
absolute volatility of the instantaneous forward rate is given by
$\sigma f_{tT} (\phi(T)-{\hat\Phi}_{tT})$.

\vspace{0.15cm}

\textbf{6. Bond option pricing}. We now turn to the problem of bond
option pricing. We consider first the price of a European-style call
option on a discount bond. Letting $t$ be the maturity and $K$ be
the strike of the option, the initial price of a bond option is
determined by the expectation
\begin{eqnarray}
C = {\mathbb E}\left[ \re^{-\int_0^t r_s {\rm d} s} \left( P_{tT}-K
\right)^+ \right].  \label{eq:20}
\end{eqnarray}
To proceed we shall apply a modification of a particular type of
change of measure technique used in Brody \textit{et al}. (2007)
for calculating option prices. Let us first examine the
denominator of the bond price appearing on the right side of
(\ref{eq:8}), and call this $\pi_t$:
\begin{eqnarray}
\pi_t = \int_t^\infty \rho_0(x)\, \re^{\sigma \phi(x) \xi_t -
\frac{1}{2}\sigma^2 \phi^2(x) t}\rd x. \label{eq:21}
\end{eqnarray}
An application of Ito's rule then gives
\begin{eqnarray}
\frac{\rd \pi_t}{\pi_t} = -r_t \rd t + \sigma {\hat\Phi}_{tt}
\rd\xi_t, \label{eq:22}
\end{eqnarray}
from which it follows, upon integration, that
\begin{eqnarray}
\pi_t = \exp\left(-\int_0^t r_s {\rm d} s + \sigma \int_0^t {\hat
\Phi}_{ss}{\rm d}\xi_s-\half\sigma^2 \int_0^t {\hat\Phi}_{ss}^2 {\rm
d} s\right) . \label{eq:23}
\end{eqnarray}
If we define further a process $\{M_t\}$ according to
\begin{eqnarray}
M_t = \exp\left(-\sigma \int_0^t {\hat\Phi}_{ss}{\rm d}\xi_s+
\half\sigma^2 \int_0^t {\hat\Phi}_{ss}^2 {\rm d} s\right) ,
\label{eq:24}
\end{eqnarray}
then a short calculation shows that the call price (\ref{eq:20}) can
be expressed in the form
\begin{eqnarray}
C = {\mathbb E}\left[ M_t \left( \int_T^\infty\! \rho_0(x)\,
\re^{\sigma \phi(x) \xi_t - \frac{1}{2}\sigma^2 \phi^2(x) t}\rd x -K
\int_t^\infty\! \rho_0(x)\, \re^{\sigma \phi(x) \xi_t - \frac{1}{2}
\sigma^2 \phi^2(x) t}\rd x \right)^+ \right]. \label{eq:25}
\end{eqnarray}
Next we substitute (\ref{eq:15}) in (\ref{eq:24}) to obtain
\begin{eqnarray}
M_t = \exp\left(-\sigma \int_0^t {\hat\Phi}_{ss}{\rm d}W_s-
\half\sigma^2 \int_0^t {\hat\Phi}_{ss}^2 {\rm d} t\right) ,
\label{eq:26}
\end{eqnarray}
which shows that $\{M_t\}$ is the change of measure density
martingale associated with (\ref{eq:15}). Letting ${\mathbb B}$
denote the `Brownian measure' under which the information process
$\{\xi_t\}$ is a standard Brownian motion we thus have
\begin{eqnarray}
C = {\mathbb E}^{\mathbb B}\left[ \left( \int_T^\infty\! \rho_0(x)\,
\re^{\sigma \phi(x) \xi_t - \frac{1}{2}\sigma^2 \phi^2(x) t}\rd x -K
\int_t^\infty\! \rho_0(x)\, \re^{\sigma \phi(x)\xi_t - \frac{1}{2}
\sigma^2 \phi^2(x) t}\rd x \right)^+ \right] \label{eq:27}
\end{eqnarray}
for the call price. Performing the Gaussian integration associated
with the ${\mathbb B}$-expectation, we obtain an explicit expression
for the price of the bond option. In particular, if $|\phi(x)|$ is
increasing then we find
\begin{eqnarray}
C = \int_T^\infty\! \rho_0(x) N\left(\sigma\sqrt{t}\phi(x)
-\frac{\xi^*}{\sqrt{t}}\right)\rd x -K \int_t^\infty\! \rho_0(x)
N\left(\sigma\sqrt{t}\phi(x)- \frac{\xi^*}{\sqrt{t}}\right) \rd x,
\label{eq:28}
\end{eqnarray}
whereas if $|\phi(x)|$ is decreasing we find
\begin{eqnarray}
C = \int_T^\infty\! \rho_0(x) N\left( \frac{\xi^*}{\sqrt{t}}-
\sigma\sqrt{t}\phi(x)\right)\rd x -K \int_t^\infty\! \rho_0(x)
N\left( \frac{\xi^*}{\sqrt{t}}- \sigma\sqrt{t}\phi(x)\right) \rd x.
\label{eq:28.5}
\end{eqnarray}
Here $\xi^*$ is the unique critical value for $\xi_t$ such that
$P_{tT}=K$. That there is a unique value for $\xi^*$ can easily be
verified by the monotonicity of the bond price in $\xi_t$. This fact
should also be intuitively clear. For example, if $|\phi(x)|$ is
increasing in $x$, then the larger the $\xi_t$ is, the more likely
that the value of $X$ is large. But if the value of $X$ is likely to
be large, then cash demand in the short time horizon is unlikely to
occur, hence resulting in higher bond prices. A converse argument
applies to the case of a decreasing $|\phi(x)|$.

\begin{figure}
\begin{center}\vspace{-0.0cm}
  \includegraphics[scale=0.75]{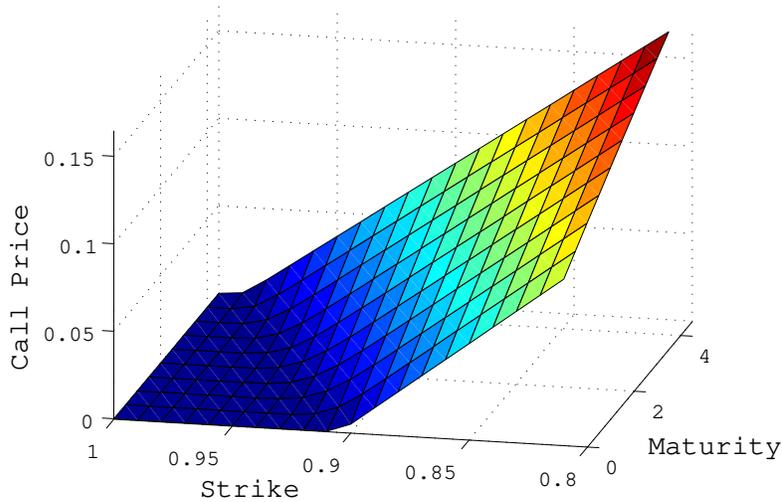}
  \vspace{-0.0cm}
  \caption{Price of a call option on a discount bond.
  The bond maturity is $5$ years. The information-adjusting
  function is set as $\phi(x)=\re^{-0.05x}$, and the initial term
  structure is assumed flat so that $P_{0T}=\re^{-0.02T}$. The
  information flow rate is set as $\sigma=0.25$.
  \label{fig:2}
  }
\end{center}
\end{figure}

Analogous calculations can be performed to obtain the price of an
option on the swap rate:
\begin{eqnarray}
C_s = {\mathbb E}^{\mathbb B}\left[ \left( \int_t^{T_n}\!
\rho_0(x)\, \re^{\sigma \phi(x) \xi_t - \frac{1}{2} \sigma^2
\phi^2(x) t}\rd x - K \sum_{i=1}^n \int_{T_i}^\infty\! \rho_0(x)\,
\re^{\sigma \phi(x) \xi_t - \frac{1}{2} \sigma^2 \phi^2(x) t}\rd x
\right)^+ \right]. \label{eq:29}
\end{eqnarray}
The point is that the random variable $\xi_t$ appearing here is
Gaussian with mean zero and variance $t$ in the ${\mathbb
B}$-measure, and hence (\ref{eq:29}) reduces to merely performing a
single Gaussian integration. Thus we see that in the present
framework we can obtain semi-analytic pricing formulae, involving
elementary Gaussian integrations, for both caplets and swaptions.

We note that the value of the parameter $\sigma$ can be calibrated
from the option price (\ref{eq:28}) or (\ref{eq:28.5}). Whether this
can always be done consistently depends on whether the option vega
defined by ${\mathcal V} = \partial C/\partial\sigma$ changes its
sign. We have considered special cases for the \textit{a priori}
density $\rho_0(x)$ and various specifications of $\phi(x)$ to
confirm the positivity/negativity of ${\mathcal V}$, suggesting that
either ${\mathcal V}>0$ or ${\mathcal V}<0$ holds for arbitrary
$\rho_0(x)$ and monotonic $\phi(x)$. Hence it seems plausible that
`implied volatility' $\sigma(K)$ in the present framework can always
be determined unambiguously from option prices.

\vspace{0.15cm}

\textbf{7. Interpretation of the auxiliary measure}. The probability
measure ${\mathbb B}$ introduced here somewhat artificially for the
purpose of calculating derivative prices in fact embodies an
economic interpretation. This can be seen from the expression
(\ref{eq:14}) for the bond price volatility $\Sigma_{tT}$, which
shows that the $\Sigma_{tT}$ is invariant under the shift $\phi(X)
\to \phi(X) + \alpha_t$ in the drift of the information process for
any $\{\alpha_t\}$ independent of $X$. This degree of freedom can be
used to fix the risk premium according to $\lambda_t = -\sigma {\hat
\Phi}_{tt}$ without loss of generality (see also the discussion in
Brody \& Hughston 2002). Thus we find that the auxiliary measure
${\mathbb B}$ introduced for the purpose of option pricing can be
identified with the market measure. It also follows that the
constant initial term structure model (\ref{eq:11}) can now be seen
as an example of the semilinear model introduced in Brody \&
Hughston (2002).

The key point to note here is that the specification of a model
$\{\xi_t\}$ for the flow of information concerning the timing of the
liquidity risk fixes both the short rate process $\{r_t\}$ and the
risk premium process $\{\lambda_t\}$. As a consequence, we find that
the process $\{\pi_t\}$ defined by the expression
\begin{eqnarray}
\pi_t = \int_t^\infty p_t(x) \rd x  \label{eq:30}
\end{eqnarray}
appearing, for example, in the denominator of the bond price
(\ref{eq:8}), is the \textit{pricing kernel}, where $p_t(x)$
denotes the unnormalised conditional probability density for the
random variable $X$ given ${\mathcal F}_t$:
\begin{eqnarray}
{\mathbb Q}\left( X\geq t | {\mathcal F}_t \right) =
\frac{\int_t^\infty p_t(x) \rd x}{\int_0^\infty p_t(x) \rd x} .
\label{eq:31}
\end{eqnarray}
In other words, $\{p_t(x)\}$ solves the Zakai equation associated
with the filtering equation for $\phi(X)$. This observation is
useful in considering the pricing of hybrid derivatives on account
of the fact that if $H_T$ represents the random payout on the
maturity day $T$ of a derivative contract (for example, the payoff
$H_T=(S_T-K)^+$ of a European option on a stock), then the price
$H_{tT}$ of the derivative at time $t\leq T$ is given by
\begin{eqnarray}
H_{tT} = \frac{{\mathbb E}^{\mathbb B}[\pi_T H_T]}{\pi_t}.
\label{eq:32}
\end{eqnarray}
In particular, for $H_T=1$ we recover the bond pricing formula.
Similarly, if $\{\pi_t^i\}$ is the pricing kernel calibrated to
currency $i$ and $\{\pi_t^j\}$ is the pricing kernel calibrated to
currency $j$, then the arbitrage-free and friction-free foreign
exchange rate is given by the ratio $\pi_t^i/\pi_t^j$. In
particular, it is the market perception of the random variables
$X^i$ and $X^j$ that determines the exchange-rate volatility (for a
pricing-kernel approach to interest rate and foreign exchange
system, see Brody \& Hughston 2004 and references cited therein).

\begin{figure}
\begin{center}\vspace{-0.0cm}
  \includegraphics[scale=0.7]{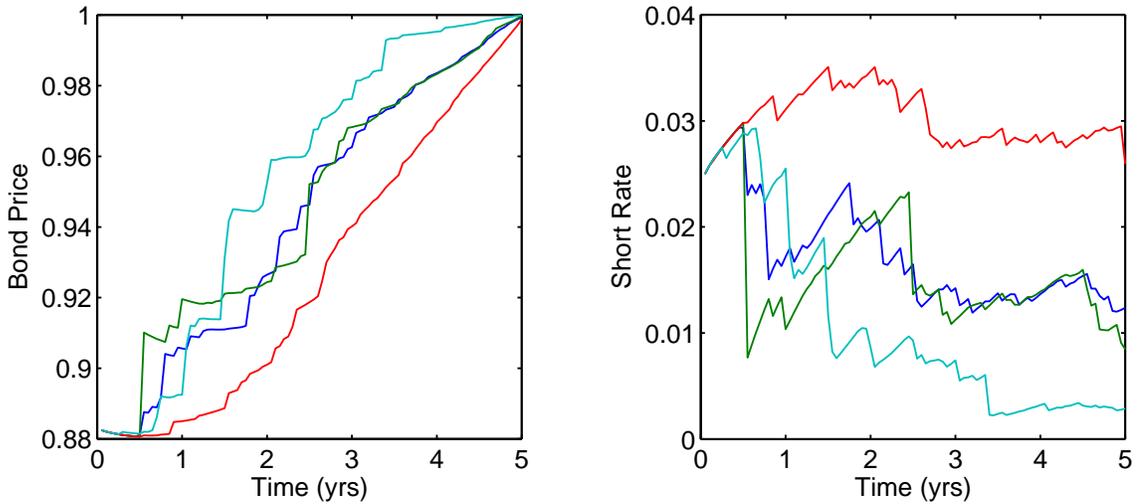}
  \vspace{-0.0cm}
  \caption{Sample paths of the discount function (\ref{eq:x36}) and
  the associated short rate (\ref{eq:x37}). The information-adjusting
  function is set as $\phi(x)=\re^{-0.02x}$, and the initial term
  structure is assumed flat so that $P_{0T}=\re^{-0.025T}$. The
  rate is set as $m=0.1$, and the bond maturity
  is 5 years.
  \label{fig:3}
  }
\end{center}
\end{figure}

It should be remarked parenthetically that although in the present
investigation we have closely examined a simple model (\ref{eq:5})
for the market information, many of the derived concepts are
applicable to a wide range of models for $\{{\mathcal F}_t\}$. Of
course, expressions for various quantities such as the bond price or
forward rate volatility are dependent on the choice of the
information process. For example, if the rate at which information
concerning the value of $\phi(X)$ is revealed to the market is time
dependent, then the Markov information model (\ref{eq:5}) will be
modified to a non-Markov process: $\xi_t = \phi(X) \int_0^t \sigma_s
\rd s + B_t$. The price of the bond in such a scenario can still be
calculated straightforwardly, with the result
\begin{eqnarray}
P_{tT} = \frac{ \int_T^\infty \rho_0(x)\, \re^{\phi(x) \int_0^t
\sigma_s {\rm d}\xi_s - \frac{1}{2}\phi^2(x) \int_0^t \sigma_s^2
{\rm d}s}\rd x} { \int_t^\infty \rho_0(x)\, \re^{\phi(x) \int_0^t
\sigma_s {\rm d}\xi_s - \frac{1}{2}\phi^2(x) \int_0^t \sigma_s^2
{\rm d}s}\rd x}. \label{eq:34}
\end{eqnarray}
Alternatively, we may consider an interest rate model driven by pure
jump L\'evy processes. An example is given by the gamma filter
introduced in Brody \textit{et al}. (2008), whereby the information
process is $\xi_t = X \gamma_t$. Here $\{\gamma_t\}$ denotes a
standard gamma process with rate parameter $m$. In this case, the
expression for the bond price process reads
\begin{eqnarray}
P_{tT} = \frac{\int_T^\infty \rho_0(x) x^{-mt} \re^{-\xi_t/x}\rd x}
{\int_t^\infty \rho_0(x) x^{-mt} \re^{-\xi_t/x}\rd x},
\label{eq:x36}
\end{eqnarray}
and the associated short rate process is
\begin{eqnarray}
r_{t} = \frac{\rho_0(t) t^{-mt} \re^{-\xi_t/t}}{\int_t^\infty
\rho_0(x) x^{-mt} \re^{-\xi_t/x}\rd x} . \label{eq:x37}
\end{eqnarray}
Hence the present framework provides for a wide range of new
interest rate models to be created that are tractable and relatively
easy to implement.

\vspace{0.15cm}

\textbf{8. Interpretation of the function $\phi(x)$}. Let us now
discuss the choice of the function $\phi(x)$ introduced in the
information process (\ref{eq:5}). What might appear to be the most
natural candidate for $\phi(x)$ is a linear function: $\phi(x)=x$.
As indicated above, in this case the bond volatility has a natural
characterisation given by the difference between (\ref{eq:17}) and
(\ref{eq:16}). We find, however, that in this model the market price
of risk (the excess rate of return above the short rate) is strictly
negative. From information-theoretic point of view, such a market
can be interpreted as follows. Recall that a small value of $X$
implies that there is an acute liquidity crisis awaiting. However,
according to the information process $\xi_t=\sigma t X + B_t$ there
is little prior warning to the market concerning the `smallness' of
$X$, since $\{\xi_t\}$ in this case is dominated by noise. As a
consequence, liquidity issues arise essentially as a surprise. What
the model shows is that in such a scenario the risk premium becomes
negative.

Under normal market conditions, we expect the risk premium be
positive. This is obtained by a function $\phi$ such that it
decreases to zero (or increases to zero if $\phi\leq0$) for large
$X$. Examples of this type are $\phi(x)=\pm\re^{-\kappa x}$ for a
positive $\kappa$, or $\phi(x)=\pm (x-x_0)^{-1}$ for a positive
$x_0$. For these choices, small values of $X$ carry heavier weights
in the signal component, as compared to larger values of $X$. As a
consequence, signals of an imminent liquidity crisis reach the
market ahead of time, enabling appropriate precautions to be taken,
thus leading to a positive risk premium. This feature can be
understood in conjunction with the fact that the observation
$\{\xi_t\}$ is a Brownian motion in the market measure. The bond
volatilities in these examples are determined by certain weighted
annuity prices. For example, when $\phi(x)=\re^{-\kappa x}$ we have
\begin{eqnarray}
\Sigma_{tT} = \left(\re^{-\kappa T} - \re^{-\kappa t}\right) -
\kappa \left[P_{tT}^{-1} \int_T^\infty \re^{-\kappa x} P_{tx}\rd x
- \int_t^\infty \re^{-\kappa x} P_{tx}\rd x\right].
\end{eqnarray}

It is worth remarking that the positivity of the risk premium in the
arbitrage pricing theory is an assumption that cannot be deduced
from the no arbitrage condition. Hence it is satisfying that in the
present framework we are able to deduce how the structure of the
flow of information affects the signature of the risk premium. In
particular, it may be possible to use the functional degree of
freedom $\phi(x)$ to calibrate various derivative prices. This is
useful, because it is virtually impossible to estimate the risk
premium directly from the price process of risky assets.

\vspace{0.15cm}

\textbf{9. Discussion}. Empirical studies indicate that a persistent
increase in money supply leads in short term (up to a month or so)
to a fall in nominal interest rates---this is the so-called
liquidity effect (Cochrane 1989). On the other hand, in the longer
term an increase in money supply increases expected inflation, hence
leading to an increase in nominal rates---this is the so-called
Fisher effect. Typically both effects coexist in that an increase in
money supply reduces nominal rates but increases expected inflation
so that the real rate also falls. Needless to say, interrelations
between these effects are difficult to disentangle. The implication
of these macroeconomic considerations to the present approach is
that the random variable $X$, which we identified as representing
the timing of liquidity crisis in the narrow sense of cash demand,
is dependant on a number of market factors and not merely on money
supply.

Going forward, we would like to formulate a model for real discount
bonds, thus allowing us to price inflation-related products in a
manner consistent with the nominal interest rate dynamics. One way
of realising this within the information-based pricing framework
might be through using multiple market factors (cf. Macrina 2006).
This would allow inflation to be modelled consistently with the
interest rate term structure.

Our objective here has been the introduction of a new interest rate
modelling framework that captures some important macroeconomic
elements, in such a way that resulting models can be used in
practice for the pricing and risk management of interest rate
derivatives. The random variable $X$, whose existence is ensured by
the positivity of nominal rates and the vanishing of
infinite-maturity bond prices, has the dimension of time, and hence
it has been interpreted as representing the timing of future
liquidity crises. It is worth emphasising that all the results
established here are, of course, independent of this particular
interpretation. Nevertheless, our interpretation allows us to
enhance fixed-income risk management with an intuitive understanding
of the model being used.


\begin{acknowledgments}
The authors thank Mark Davis, Lane Hughston, Bernhard Meister,
Mihail Zervos, and in particular, Martijn Pistorius for stimulating
discussions.
\end{acknowledgments}

\end{document}